\documentclass[conference]{IEEEtran}
\IEEEoverridecommandlockouts

\usepackage{cite}
\usepackage{amsmath,amssymb,amsfonts}
\usepackage{algorithmic}
\usepackage{graphicx}
\usepackage{textcomp}
\usepackage{xcolor}
\usepackage{hyperref}
\def\BibTeX{{\rm B\kern-.05em{\sc i\kern-.025em b}\kern-.08em
    T\kern-.1667em\lower.7ex\hbox{E}\kern-.125emX}}

\usepackage{balance}

\begin{document}

\title{DNA sequence alignment: An assignment for OpenMP, MPI, and CUDA/OpenCL\\
\thanks{This work has been developed in the context of the GAMUVa group 
(\url{https://gamuva.infor.uva.es/}), and it has been partially supported by
Vicerrectorado de Innovación Docente y Transformación Digital de la Universidad de Valladolid,
Proyectos de Innovación Docente PID2324\_84, and by the NVIDIA Hardware Grant Program 
for providing GPU devices
used during the assignments.}
}

\author{\IEEEauthorblockN{Arturo Gonzalez-Escribano}
\IEEEauthorblockA{\textit{Departamento de Informática} \\
\textit{Universidad de Valladolid}\\
arturo@infor.uva.es}
\and
\IEEEauthorblockN{Diego García-Álvarez}
\IEEEauthorblockA{\textit{Departamento de Informática} \\
\textit{Universidad de Valladolid}\\
dieggar@infor.uva.es}
\and
\IEEEauthorblockN{Jesús Cámara}
\IEEEauthorblockA{\textit{Departamento de Informática} \\
\textit{Universidad de Valladolid}\\
jesus.camara@infor.uva.es}
}

\maketitle

\begin{abstract}
We present an assignment for a full Parallel Computing course. Since 2017/2018, we have proposed a different problem each academic year to illustrate various methodologies for approaching the same computational problem using different parallel programming models. They are designed to be parallelized using shared-memory programming with OpenMP, distributed-memory programming with MPI, and GPU programming with CUDA or OpenCL. The problem chosen for this year implements a brute-force solution for exact DNA sequence alignment of multiple patterns. The program searches for exact coincidences of multiple nucleotide strings in a long DNA sequence. The sequential implementation is designed to be clear and understandable to students while offering many opportunities for parallelization and optimization.
This assignment addresses key concepts many students find difficult to apply in practical scenarios: race conditions, reductions, collective operations, and point-to-point communications. It also covers the problem of parallel generation of pseudo-random sequences and strategies to notify and stop speculative computations when matches are found. This assignment serves as an exercise that reinforces basic knowledge and prepares students for more complex parallel computing concepts and structures. It has been successfully implemented as a practical assignment in a Parallel Computing course in the third year of a Computer Engineering degree program. Supporting materials for this and previous assignments in this series are publicly available.
\end{abstract}

%

\section{Idea and context}

\paragraph{Idea}
Program parallelization requires different approaches depending on the programming model used. Understanding these variations is essential for students to explore advanced techniques and effectively address the challenges of parallel programming on modern heterogeneous platforms. At the University of Valladolid, we offer a Parallel Programming elective course in the third year of the Computer Engineering degree. This course covers fundamental concepts of the shared-memory model with OpenMP, the distributed-memory model with MPI, and programming GPUs with CUDA or OpenCL. Assignments proposed in previous academic years~\cite{GAMUVa_peachyAssignments} have proven effective in teaching each programming model and providing valuable information on the portability of concepts and techniques across these models. By engaging in these assignments, students gain a deeper understanding of the similarities and conceptual shifts between different approaches, enabling them to critically analyze and select the most appropriate programming models and solutions for different problems.

\begin{figure}[t]
  \includegraphics[width=0.5\columnwidth]{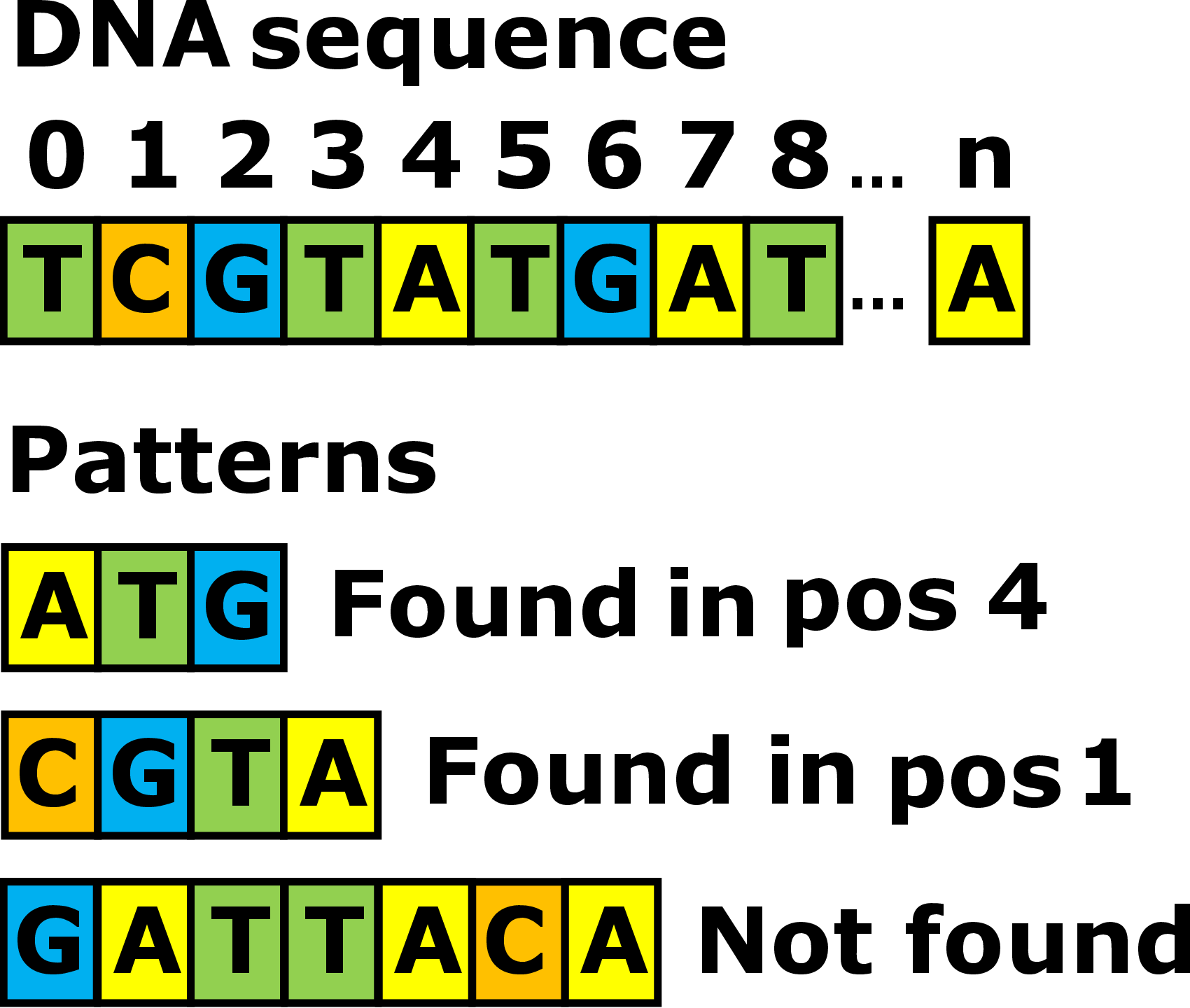}
  \caption{Example of DNA sequence alignment}
  \label{fig:figure}
\end{figure}

\paragraph{The assignment problem}
DNA sequences can be represented as strings of characters, where each character represents a nucleotide type. The \emph{exact DNA sequence alignment} problem, or \emph{pattern searching} problem, determines the exact coincidences of nucleotide strings in a long DNA sequence. Figure\ref{fig:figure} illustrates an example of multiple pattern matching within a DNA sequence. Our assignment solves this problem for a random main DNA sequence and multiple nucleotide patterns that can also be random sequences and/or exact copies of parts of the main sequence randomly located. 
Several sophisticated algorithms have been proposed for this problem (see e.g.~\cite{RehmanEtAl24}).
We choose a simple brute-force algorithm that checks each pattern starting at each possible 
position of the DNA sequence. It presents direct parallelization opportunities at two levels, patterns and starting positions. In our solution, if a pattern appears multiple times only the position of the first match is registered. In the sequential program, a pattern search stops when a match is found, skipping the search in the rest of the starting positions. The program determines which patterns are found in the sequence and their starting positions.

\section{Concepts covered}

\paragraph{Concepts}
All the assignments in this series try to cover most of the concepts taught during the course. This particular assignment covers several main concepts. The students tackle loop parallelization with two nested loops that can be parallelized independently or together, considering which variables should be shared or private. This forces to detect and solve both write and update race conditions when managing both counters for the number of matchings and an ancillary shared array to control the number of matchings on each position of the main sequence. They can be solved with critical regions, atomics, or reduction substitutions. Fingerprint checksums of the starting positions and the number of positions where more than one pattern matches are computed to help in checking the correctness when the code is modified to parallelize and optimize it. These computations also introduce non-trivial reduction operations with multiple solutions. The students should decide and test in which situations atomic operations or reductions are more profitable. The solutions to all these problems should be ported between OpenMP and CUDA/OpenCL. In CUDA/OpenCL students should first understand how to work with thread-blocks and coalesced memory accesses.
In MPI the students should distribute the main DNA sequence across processes, forcing some of the searches to start in one processor and finish in another. This leads to the creation of a pipeline pattern with point-to-point communications. Other basic collective operations and communications are needed to compute the global checksums. The program arguments can be used to generate patterns with randomly distributed lengths and locations, providing means and deviations. This allows the generation of different load distributions that can be addressed with proper load-balancing policies on each programming model targeted. The code presents opportunities for further optimizations based on cache effects, manual inlining, code elimination or simplification, etc. This assignment uses one-dimensional arrays to store the DNA sequence and patterns, to help the students avoid dynamic or complex memory management, focusing on the parallelization and optimization issues. Random DNA sequences and patterns also introduce the problem of parallelizing the generation of pseudo-random number sequences. To simplify this task, the program leverages a custom linear congruential random generator with a skip-ahead function based on PCG~\cite{oneill:pcg2014}. Looking for the first starting position where a pattern matches introduces the possibility to exploit strategies to notify and stop other speculative computations that are checking the same pattern in higher starting positions.

\paragraph{Variants}
The teacher may consider the inclusion or not of the initialization of the DNA sequence and patterns in the code targeted by the students. In MPI, the assignment can be easily simplified moving the generation of the main DNA sequence before the code targeted by the students, obtaining a duplicate of the whole sequence in all processes.

\section{Using the Assignment}

\paragraph{Teaching context}
In previous courses, students get acquainted with operating systems, concurrency concepts, and the C programming language. This year the course enrolled a total of 44 students, who worked in pairs for the assignments. 40 students attended and participated in all the activities.
In an initial session, we explain to the students the DNA sequence alignment problem and our implementation using practical examples. They are provided with a handout and the starting sequential code, where the parts and functions of the program that can be modified are clearly marked. We also provide them examples of program arguments that generate different types of main sequences and patterns, with different lengths and locations. 
For each programming model, the students attend three weeks of lectures and lab sessions 
to gradually train them in the use of the main concepts. Then, each pair of students works for one week to solve the assignment in the corresponding programming model. After this period, there is an exam with questions about how they solved specific problems in the submitted solutions, to demonstrate their understanding and commitment to the assignment.

\paragraph{Tools} 
The only software required is a modern OpenMP-compatible C compiler, any MPI library, and a CUDA or OpenCL toolkit. The OpenMP or MPI solutions can be developed and tested on any multicore computer, while a GPU board is required for the third task. Nevertheless, the best experience is obtained with a shared cluster which also allows students to compare their results in terms of performance improvement. In our course, the three one-week periods dedicated to the development of the assignment on each programming model are organized as programming contests~\cite{fresno2017gamification}. We use an online judge that evaluates and classifies the solutions in terms of correctness and performance. It also works as a queue manager to submit the students' programs to the servers of our research cluster. See the reproducibility appendix below for more details about the cluster.

\paragraph{Student's satisfaction}
The students are offered to fill out a survey at the end of the course. 26 students did it. To the question: ``Are you satisfied with the overall experience of the course, activity types, evaluation method, etc.?'', using a Likert scale from 1 to 5, the average is 3.77 and the median is 4. Some students complain about the practical work exams and their weight in the final grade. Nevertheless, they agree that the assignment illustrates the main concepts of the course and provides opportunities to go deep into the subject. For example, some students optimized their CUDA codes to execute 9.66 times faster than the baseline parallel version. The execution times obtained by the submissions, the results of the contests, and other statistical data, are publicly available at \url{http://frontendv.infor.uva.es}.  %
The OpenMP task is considered by the students the easiest one because they do not need to understand what the code is doing to get the first parallel versions. The MPI task is reported as the most difficult due to the code changes needed. The CUDA task is the most satisfying, as porting the OpenMP solutions for problems such as race conditions is quite direct, and the performance results are the most remarkable. There have been 8,956 submissions to the cluster queues, with an average of more than 446 submissions per pair of students. Around 1,400 submissions correspond to compilation errors. 4,783 submissions finished with correct results, and their execution times were considered for the student's position in the leaderboards.


\bibliographystyle{IEEEtran}
\bibliography{bibliography}

\begin{thebibliography}{1}
\providecommand{\url}[1]{#1}
\csname url@samestyle\endcsname
\providecommand{\newblock}{\relax}
\providecommand{\bibinfo}[2]{#2}
\providecommand{\BIBentrySTDinterwordspacing}{\spaceskip=0pt\relax}
\providecommand{\BIBentryALTinterwordstretchfactor}{4}
\providecommand{\BIBentryALTinterwordspacing}{\spaceskip=\fontdimen2\font plus
\BIBentryALTinterwordstretchfactor\fontdimen3\font minus
  \fontdimen4\font\relax}
\providecommand{\BIBforeignlanguage}[2]{{%
\expandafter\ifx\csname l@#1\endcsname\relax
\typeout{** WARNING: IEEEtran.bst: No hyphenation pattern has been}%
\typeout{** loaded for the language `#1'. Using the pattern for}%
\typeout{** the default language instead.}%
\else
\language=\csname l@#1\endcsname
\fi
#2}}
\providecommand{\BIBdecl}{\relax}
\BIBdecl

\bibitem{GAMUVa_peachyAssignments}
{GAMUVa group}, ``\BIBforeignlanguage{eng}{Peachy parallel assignments},''
  2018-2024, on https://gamuva.infor.uva.es/peachy-assignments/.

\bibitem{RehmanEtAl24}
S.~Ur~Rehman and {et.al.}, ``\BIBforeignlanguage{eng, es}{Smart exact string
  matching algorithm specifically for dna sequencing},'' in
  \emph{\BIBforeignlanguage{eng, es}{2024 2nd International Conference on Cyber
  Resilience (ICCR)}}.\hskip 1em plus 0.5em minus 0.4em\relax Dubai, United
  Arab Emirates: IEEE, 2024.

\bibitem{oneill:pcg2014}
M.~E. O'Neill, ``{PCG:} a family of simple fast space-efficient statistically
  good algorithms for random number generation,'' Harvey Mudd College,
  Claremont, CA, Tech. Rep. HMC-CS-2014-0905, Sep. 2014.

\bibitem{fresno2017gamification}
J.~Fresno, A.~Ortega-Arranz, H.~Ortega-Arranz, A.~Gonzalez-Escribano, and D.~R.
  Llanos, ``Applying gamification in a parallel programming course,'' in
  \emph{Gamification-Based E-Learning Strategies for Computer Programming
  Education}, R.~A.~P. de~Queirós and M.~T. Pinto, Eds.\hskip 1em plus 0.5em
  minus 0.4em\relax IGI Global, 2017, ch.~6, pp. 106--130.

\end{thebibliography}

\section*{Appendix: Artifact description and Reproducibility}
\balance

The assignment has been used in the context of a Parallel Computing course, in the third year of the Computing Engineering grade at the University of Valladolid (Spain).

The material of the assignment, including a handout, the starting sequential code, and some program arguments to be used as examples are publicly available through the CDER courseware repository and our Peachy Assignments web page: \url{https://gamuva.infor.uva.es/peachy-assignments/}.

The online judge program used in the programming contests is named \emph{Tablon}. It was developed by the Trasgo and GAMUVa research and education innovation groups at the University of Valladolid (\url{https://trasgo.infor.uva.es/tablon/}). It uses the Slurm queue-management software to interact with the machines in the cluster of our research group. During the course, we used the Slurm 23.11.6 release.

The machine of the cluster used for the OpenMP contest is named \emph{heracles}. It is a server with four AMD Opteron 6376 @ 2.3Ghz CPUs, with 64 cores, and 128 GB of RAM. It shows interesting effects related to memory management and memory accesses due to its 4 NUMA nodes. 

The machines used for the CUDA/OpenCL contests are named \emph{gorgon} and \emph{medusa}. Gorgon is a server with two AMD EPYC 7713 64-Core Processor @2.0 GHz CPUs, and 512 GB of RAM, with 128 physical cores using SMP to provide 256 threads. It is equipped with 1 NVIDIA A100 and 2 NVIDIA's RTX4500 GPUs.  Medusa is a server with two Intel Xeon Silver 4208 CPUs @2.10GHz, with 16 cores with hyperthreading (32 threads) @1.4Ghz. It is equipped with	1 NVIDIA TitanX GPU.

During the MPI contest, we use \emph{medusa} and \emph{gorgon} together. They are interconnected by a 40Gb Ethernet network fabric. All machines are managed by a Rocky 9 operating system. The compilers and system software used are GCC v11.4, and CUDA v12.4.

The results of the contests, execution times, status of all submissions, and other statistical data, for the last six peachy assignments in this series, are publicly available at \url{http://frontendv.infor.uva.es}.


\end{document}